# Negative Differential Resistance in Carbon-Based Nanostructures

S.A. Evlashin[a], M.A. Tarkhov[b], D.A. Chernodubov[c], A.V. Inyushkin[c], A.A. Pilevsky[d], P.V. Dyakonov[a,d], A.A. Pavlov[b], N.V. Suetin[d], I.S. Akhatov[a], V. Perebeinos[e*]

[a]*Center for Design, Manufacturing & Materials, Skolkovo Institute of Science and Technology, 121205, Russia*

[b] *Institute of Nanotechnology of Microelectronics of the Russian Academy of Sciences, Moscow, 119991, Russia*

[c]*National Research Center "Kurchatov Institute," Moscow 123182, Russia*

[d]*Skobel'tsyn Institute of Nuclear Physics, Lomonosov Moscow State University, Moscow, 119991, Russia*

[e] *Department of Electrical Engineering, University at Buffalo, The State University of New York, Buffalo, New York 14260, USA*

* vasilipe@buffalo.edu

Nonlinear electrical properties, such as negative differential resistance (NDR), are essential in numerous electrical circuits, including memristors. Several physical origins have been proposed to lead to the NDR phenomena in semiconductor devices in the last more than half a century. Here, we report NDR behavior formation in randomly oriented graphene-like nanostructures up to 37 K and high on-current density up to $10^5$ A/cm$^2$. Our modeling of the current-voltage characteristics, including the self-heating effects, suggests that strong temperature dependence of the low-bias resistance is responsible for the nonlinear electrical behavior. Our findings open opportunities for the practical realization of the on-demand NDR behavior in nanostructures of 2D and 3D material-based devices via heat management in the conducting films and the underlying substrates.

## I. Introduction

Nonlinear electrical characteristics allow the design of devices with unique properties. In particular, the realization of such components and devices as electronic circuitry, storage memory, electric switches, wave generators, power generators, and multipliers [1–4]. Most commonly, two-terminal Schottky diodes, Esaki diodes, and Gunn diodes [5,6] exhibit nonlinear electrical characteristics. Three-terminal field-effect transistors with nonlinear characteristics have also been demonstrated [7–9]. The interest in nonlinear electrical devices was revived after an experimental demonstration of a memristor by the HP lab [10]. Recently, devices with highly nonlinear current-voltage characteristics found applications as selectors in crossbar memory arrays [11,12]. However, the significant obstacles in applications include the poor on-current density, integration challenges, and manufacturing cost.

Unlike "Ohmic" resistances, negative differential resistance (NDR) demonstrates an increase of current at a decrease of voltage and vice versa. NDR is of particular interest, among the other nonlinear properties. Depending on the I-V curve shape, NDR can be of N-type voltage control or S-type current control [13]. Recently, NDR devices have been realized on different oxide films [14–20], semiconductor heterostructures [21,22], single molecules [1,23,24], 2D materials, and their combinations [3,25–30]. The origins of the NDR behavior can be due to the band structure effects, phonon blockade, chemical reactions, heat management, electrical contacts, to name a few [1-30].

NDR behavior in metal oxide films has attracted much interest [31–33]. The S-type NDR in such systems is associated with the formation of current density domains parallel to the electric field [32] and Poole-Frenkel effects [31]. Poole-Frenkel model describes the NDR in the $NbO_x$ films [31] by the presence of donors and traps. In TiO films, the NDR was explained by the Joule heating and polaronic transport [32]. In the recent theoretical work [35], the authors demonstrated the influence of the Joule heating on NDR and compared their results with the earlier findings [2,36]. They concluded that strong temperature dependence of the low-bias resistance leads to the onset of NDR. Therefore, in principle, NDR can result from any transport mechanism as long as thermal runaway occurs in sustainable fields and temperatures.

Carbon nanostructures can also demonstrate nonlinear characteristics. NDR was reported in mono- and multilayered graphene films, defective and n-doped graphene, and graphene oxide, carbon nanotubes [3,9,29,30,37–42]. The NDR in the defected or doped graphene layer is attributed to a change in the band structure [29,30]. In graphene nanoribbons, it was attributed to the interference at zigzag edges [43]. It is well known that carbon has a narrow electrochemical window and can be easily oxidized [44]. In Ref. [39], the authors demonstrated NDR in graphene oxide by applying different current densities in the graphene oxide film leading to oxidation and reduction of these films.

This work demonstrates a nanostructure class of materials that possess S-type NDR characteristics and high current density capacity. These materials consist of sponge-like carbon-based material called carbon NanoWalls (CNWs). Our simulations demonstrate that the direct influence of the Joule losses causes NDR phenomena, and we suggest that thermal management between the film and the substrate interface is essential for the NDR characteristics.

## II. Experimental Methods

*Synthesis of Carbon NanoWalls*

CNWs are synthesized on a silicon substrate by Direct Current Plasma Enhanced Chemical Vapor Deposition (DC PECVD) method [45,46]. A mixture of $CH_4$ and $H_2$ flows are 15 sccm and 200 sccm, correspondingly, was used as a source gas. The pressure is 150 Torr, and the discharge current is 5.2 A, and synthesis time is 20 min for the thin sample and 60 min for the thick sample.

*Structure characterization*

The obtained structures are characterized by Carl Zeiss Supra 40 Scanning Electron Microscopy (SEM) and Raman spectroscopy (Thermo Scientific DXR Raman Microscope spectrometer). For more details, see [47,48].

*Fabrication of meander*

The laser ablation technique is applied for meander fabrication [49]. Pulse laser TEMA with 1064 nm wavelength, 70 fs pulse duration, laser power 0.1 W with pulse reputation ratio 80 MHz. The meander area is 2 x 2 mm, the line width of 100 µm with a gap between contacts of 100 µm.

*Current-voltage measurements*

The resistivity measurements *R(T)* are conducted in Sumitomo Heavy Industries SRDK 101D cryocooler. The current-voltage characteristics are measured using a low-noise precision current source Keithley Instruments 2400. Temperature monitoring is conducted by a Lake Shore Cryotronics DT-670-SD temperature controller. This system allows temperature control with a rate of heating/cooling of 1 K/min. The stability range is from 2.3 to 230 K. The current-voltage characteristics are measured at a fixed temperature in the current or voltage stabilization regime. The pressure at the chamber was $10^{-7}$ Torr.

*Thermal conductivity measurements*

The thermal conductivity is measured using a steady-state longitudinal heat flow method in the temperature range from 5 to 100 K. One end of the sample is fixed in a copper block that serves as a heat sink, while a heater is mounted to another end. The temperature gradient is defined as a ratio of the temperature difference between the two points of the sample located along its long axis. The distance between these points is about ~ 3.5 mm. The heat output in the heater is selected so that the relative temperature difference is approximately 0.01. In order to minimize the heat losses from the heater, the measurements are performed in a vacuum cell with a residual gas pressure of less than $7x10^{-6}$ Torr. The technique used was similar to that described in Ref. [50] with an exception: in this work, the temperature gradient is measured using a pair of Cernox CX-1050 resistance temperature sensors (Lake Shore Cryotronics, Inc.), while in Ref. [50] the thermocouples were utilized. The thermal conductivity of the film is found from the difference in the thermal conductivities of the substrate measured with and without the film. During the measurements, the film from the quartz substrate is removed mechanically in situ, without demounting the sample from the measuring cell.

## III. Results

An SEM image of the thin film of Carbon NanoWalls is shown in Fig. 1a, and a schematic representation of the film morphology is shown in Fig. 1b. The thick film has the same structure, except for the thickness is about one micron. The corresponding Raman spectra for thin and thick films are shown in Fig. 1(c). A defect ratio of the carbon structures depends on the synthesis parameters. The intensity ratio of the D to G peaks changes from 1.27 to 1.10, which corresponds to a change in an average crystallite size from 15.0 to 17.4 nm [51,52]. Similar dependencies are observed for the intensity ratio of 2D to G peaks, which changes from 1.22 to 1.11.

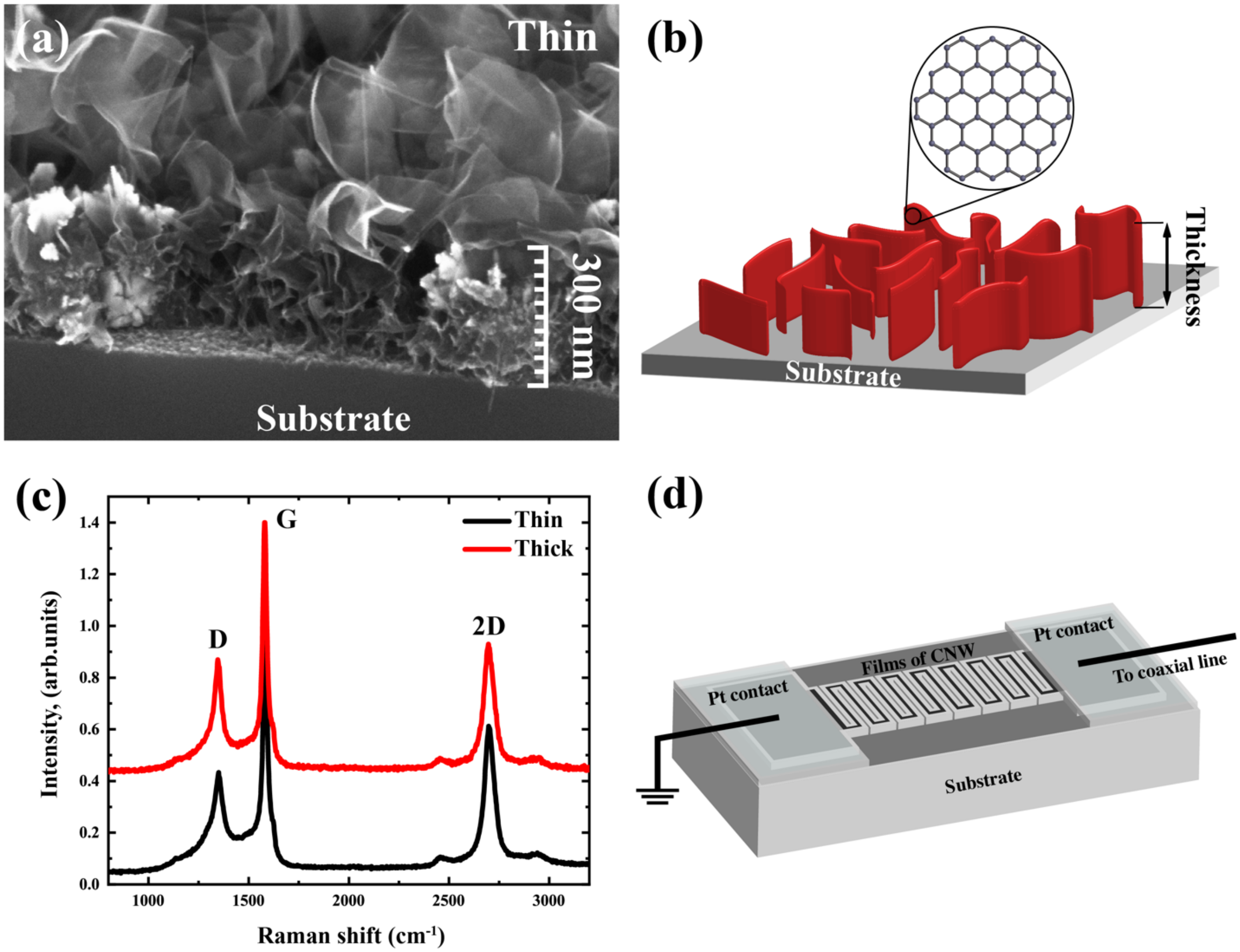

Fig. 1. (a) SEM images of CNWs synthesized for 20 min (b) the synthesis time is 60 min, (c) Raman spectra of samples of various synthesis times. (d) a schematic representation of the measuring device.

A schematic image of the electrical device is shown in Fig. 1(d). The current-voltage characteristic (I-V curve) for the thin film demonstrates a negative differential resistance at temperatures lower than 37 K, as shown in Fig. 2(a). Note that the current density in our device operating in NDR regime reaches values of up to $10^5$ A/cm$^2$. At the same time, the thick film of 1000 nm thickness does not show such behavior. The measured I-V curves on the thick films show Ohmic dependencies and are not shown in the manuscript.

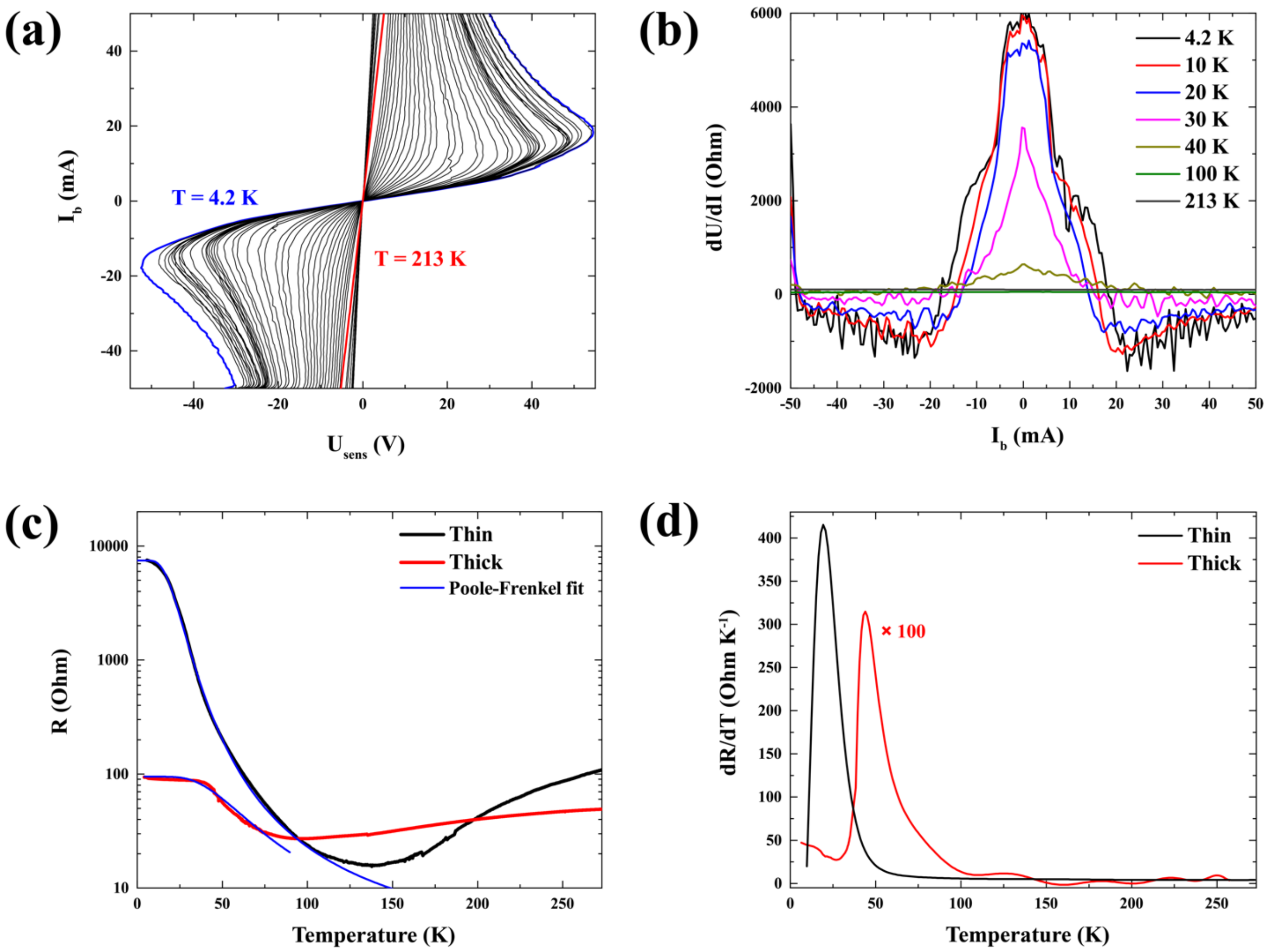

Fig. 2. (a) The current-voltage characteristics at different ambient temperatures for thin CNW film, (b) differential resistance at different temperatures thin films of CNWs. (c) temperature dependencies of the low-bias resistance along with the fits to the Poole-Frankel model Eq. (1), (d) the corresponding temperature dependencies of the low-bias differential resistance *dR/dT*. Note a factor of 100 used for the red curve.

Fig. 2(b) demonstrates the temperature and bias regime when NDR is most pronounced. Fig. 2(c) shows the temperature dependencies of the low-bias electrical resistance. A similar shape of dependencies is observed for CNWs by other researchers [53]. For the thin film, the resistance reaches its minimum value at 130 K, and the differential resistance *dR/dT* shows a peak at around 21 K, as shown in Fig. 2(d) by the black curve. The thick film demonstrates the minimal resistance at 90 K, and the maximum differential resistance at a much smaller value than the thin film, as shown in Fig. 2(d) by the red curve.

Previous studies have shown that a considerable temperature gradient is responsible for the formation of S-type NDR in oxide films [18]. The formation of high-current density domains and thermally activated Poole-Frankel conductivity, together with the Joule heating, lead to the NDR behavior in oxide films [18]. This Joule heating mechanism is not limited to metal oxides and can be extended to other materials. In our samples, the low-bias resistance at low temperatures can well be accounted for by the Poole-Frankel model [18]:

$$R(T) = \frac{R(0)}{((c*e^{\left(-\frac{T_t}{T}\right)})^n+1)^{\frac{1}{n}}} \quad (1)$$

The fitting parameters *R(0),* $T_t$, *c*, and n are given in Table 1. For both films, the Poole-Frankel model describes the data remarkably well at temperatures below 100 K, as shown in Fig. 2(c).

Table 1. Fitting parameters for the Poole-Frankel model in Eq. (1).

| | **Thickness, nm** | **R(0) (Ohm)** | **$T_t$ (K)** | **c** | **n** |
|---|---|---|---|---|---|
| **Thin** | 400 | 7470 | 360 | 1790 | 0.2 |
| **Thick** | 1000 | 94.7 | 360 | 61.8 | 0.474 |

The thin film's activation energy is found to be 31 meV or $T_t$=360 K. We do not observe considerable variation in the low-bias resistance in the thick film to reliably extract the activation energy. Since we expect the nature of the traps to be similar in both films, we fixed the parameter $T_t$ at the same value. It is worth mentioning that the room temperature low-bias electrical conductivity of the CNW films is of the order of $10^6$ S/m, which is much higher than that in typical NDR devices.

**Discussion**

To simulate I-V characteristics at higher biases, we solve the self-consistently Fourier equation for the heat flow with the transport Ohm's law equation. The temperature variation along the channel of length *L*=40 mm is given by:

$$\frac{I^2R(T(x))}{WL} + \kappa_{CNW}(T(x))h_{CNW}\,\frac{d^2T(x)}{dx^2} - Q_{sub} = 0,$$

$$T(x=0) = T(x=L) = T_0 \quad (2)$$

Where $T_0$ is the ambient temperature, and *R(T)* is the low-bias resistance from Fig. 2(c). The channel length is 40 mm and the applied bias of 50 V would translate to the electric field of about 0.001 V/µm. Therefore, our devices are still in the low-bias regime operation. The first term in Eq. (2) gives the Joule losses. The current density is constant according to the continuity equation. The second term in Eq. (2) gives the in-plane heat dissipation across the film. We assume a constant temperature profile across the film.

The measured temperature dependence of the film's thermal conductivity $\kappa_{CNW}$ is shown in Fig. 3(a). The thermal conductivity of CNW film $\kappa_{CNW}$ plays a critical role in establishing temperature distribution along the channel. The simulations of the temperature distribution along the channel are shown in Fig. 3(b). We find that temperature reaches its maximum value on a length scale of $L_c$=15-30 µm from the metal contacts in our devices. These results demonstrate the potential of scaling of the channel lengths in these devices. Below $L_c$, NDR would be challenging to achieve due to the metal electrodes' efficient heat sink.

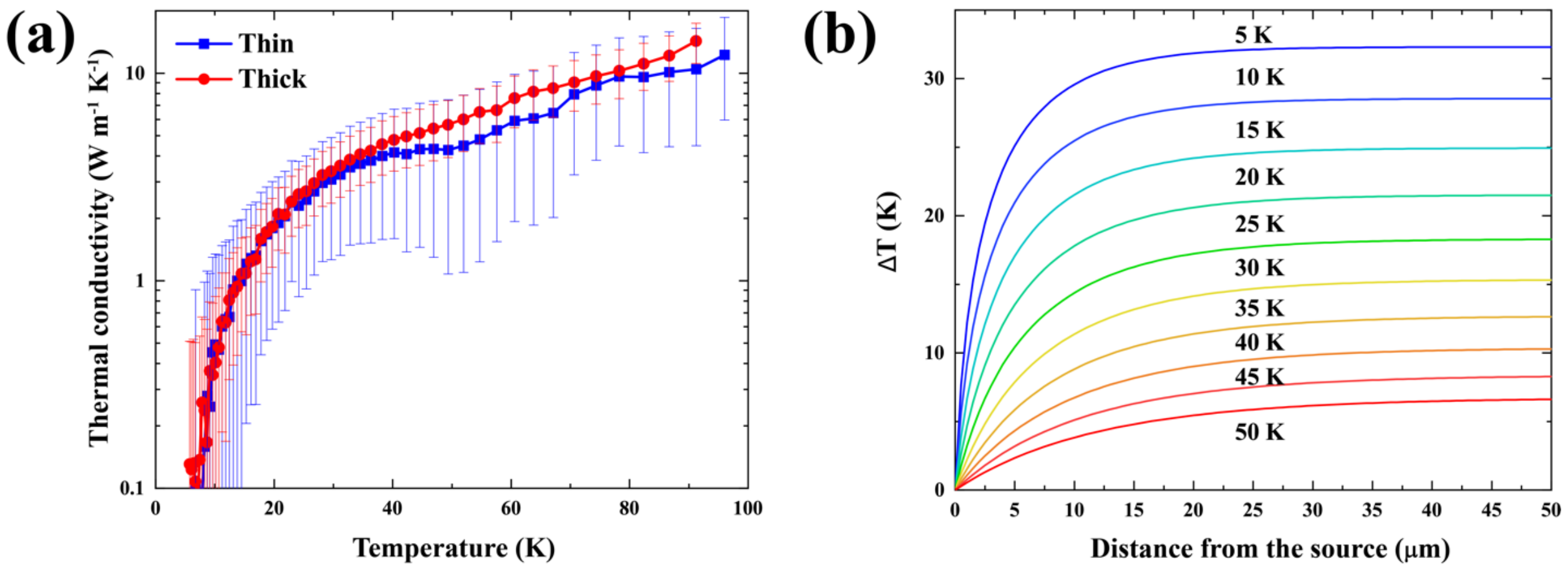

Fig. 3. a) The measured thermal conductivity of CNW film $\kappa_{CNW}$ by the steady-state longitudinal heat flow method, b) simulated temperature variation along the channel near the metal contact. The temperature reaches its maximum on a length scale of a few tenths of microns.

The last term in Eq. (2) gives thermal dissipation to the substrate:

$$Q_{sub} = \kappa(T(x) - T_0), \quad (3)$$

where $\kappa$ is the effective thermal conductance, which includes both interfacial thermal conductance between the CNW film and the substrate and the thermal conductance of the substrate itself. The latter can be estimated as $\kappa_{ox}/t_{ox}$, where $\kappa_{ox}$ is the silicon substrate thermal conductivity of thickness $t_{ox}$. The film cross-section can be found from the $W$=100 μm and thickness $h_{CNW}$=0.4 μm or 1.0 μm for the thin and thick films, correspondingly.

Once the temperature profile is calculated from Eq. (2) and (3), the Ohm's law gives the corresponding voltage drop across the channel: $V = \frac{I}{L}\int_0^L R\big(T(x)\big)\,dx$, as a function of current *I*.

The results of I-V simulations for the thin CNW film are shown in Fig. 4(a). Our simulations reproduce experimental results in Fig. 2(a) remarkably well and demonstrate that NDR behavior is caused by self-heating. The corresponding maximum temperatures in the channel as a function of current are shown in Fig. 4(b) for devices at different ambient temperatures. On the other hand, simulations in Fig. 4(c) with the experimental low-bias *R(T)* from Fig. 2(c) for the thick CNW film do not demonstrate NDR behavior consistent with the experiment (not shown). The main reason for the linear I-V characteristics in the latter case is the much weaker temperature dependence of the low-bias resistance.

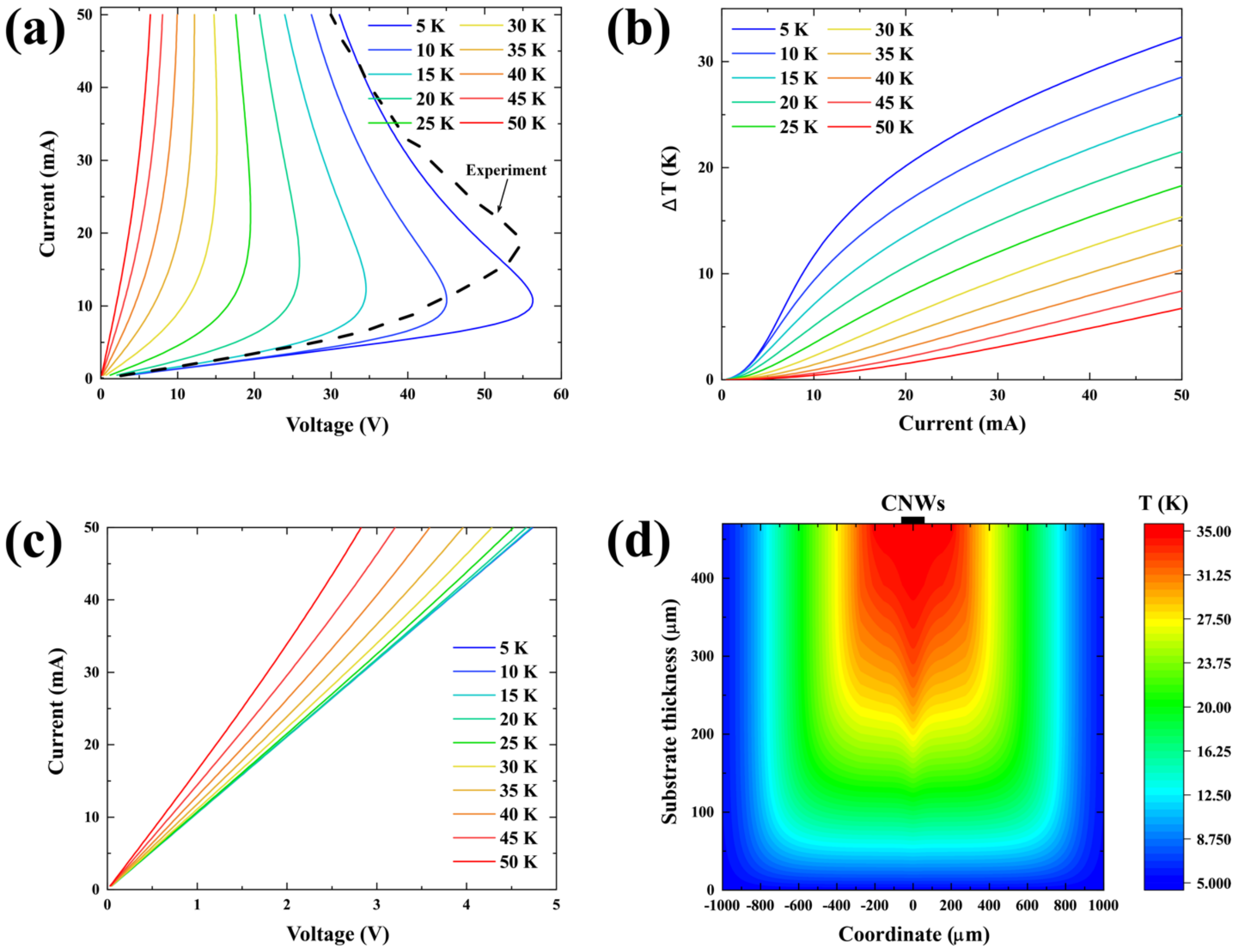

Fig. 4. (a) Simulated I-V curves for the thin film at ambient temperatures from 5K to 50K with a step of 5K: from top to bottom, using $\kappa=12000$ W/Km$^2$ a single fitting parameter. The dashed curve shows experimental I-V with the most pronounced NDR. While the overall agreement is rather good, some deviations are likely due to the temperature dependence of $\kappa$, which is not considered in our modeling. (b) The maximum temperature in the channel relative to the ambient temperature for the same conditions as in simulation in panel a. (c) Simulated I-V curves for the thick film at ambient temperatures from 5K to 50K with a step of 5K: from top to bottom, using the same value of $\kappa$ and the experimental low-bias resistance *R(T)* from Fig. 2c, (d) 2D simulation of the heat dissipation to the substrate, which is sitting of a stage at 5K. The temperature of the CNW film, shown by the black bar on top, is 35K.

Table 2. NDR characteristics of different material classes.

| | **Materials** | **NDR type, mechanism** | **Current density, A/cm$^2$** | **Peak-to-Valley Ratio** | **Temperature** | **Ref.** |
|---|---|---|---|---|---|---|
| **Semiconductors** | *Si based devices* | N, tunneling | ~1000 | ~2-4 | 300 K | [54] |
| | *Ge based devices* | N, tunneling | ~1000-16000 | ~7-10 | 300 K | [54] |
| | *III-V based devices* | N, tunneling | ~1300 | ~20 | 300 K | [54] |
| | *SiGe/Si* | N, tunneling | ~0.2 | 7.6 | 300 K | [55] |
| | GaN/AlN | N, tunneling | ~7000 | 2.2 | 300 K | [21] |
| | Si doped with Sb | S, Joule-heating | 0.1 | 1.3 | 1.9 K-2.5 K | [56] |
| | *$CaF_2/CdF_2$* | N, tunneling | 63 | ~$10^5$ | 300 K | [57] |
| **Organic molecules** | Molecules | N, tunneling | $2.5 \times 10^5$ (25 nA per molecule)[a] | 15 | 6 K | [7,24] |
| | Molecules | N, tunneling | $1.5 \times 10^4$ (1.5 nA per molecule)[a] | ~10 | 300 K | [7,23] |
| **Oxides** | $VO_2$ | N, electric-field induced resistance switching | ~40000 | 2.3 | 300 K | [14] |
| | $VO_2$ | S, Joule-heating driven thermal phase transition | ~$1.7 \times 10^5$ | ~1.5 | 300 K | [58] |
| | $TiO_2$ | N, charge trapping/detrapping | ~2000 | 1.7 | 300 K | [17] |
| | $TiO_2$ | N, electrochemical reactions | ~0.15 | ~1.5 | 300 K | [59] |
| | $TaO_x$ | S, Joule-heating | ~1000 | ~3 | 300 K | [18, 59] |
| | $Nb_2O_5/NbO_x$ | S, Joule-heating | ~15 | 1.3 | 233 K-403 K | [60] |
| | $NbO_x$ | S, Joule-heating | 5 x $10^6$ | 1.6 | 275 K - 450 K | [31] |
| | $BaTiO_3$ | N, charge trapping / detrapping | ~2 | 1.5 | 300 K | [15] |
| **Layered materials** | Black Phosporene/InSe | N, tunneling | ~0.1 | 3 | 10-100 K | [22] |
| | Graphene | N, electron-hole asymmetry in ambipolar regime | ~$10^9$ (~2 mA/µm)$^b$ | 1.1 (~2.5) | 5 K (300 K) | [9] |
| | Graphene nanoribbon | N, tunneling | ~$10^7$ (~32 mA/mm)$^b$ | 5.9 | 300 K | [61] |
| | graphene oxide | N, electrochemical reactions | ~$10^5$ (~0.2 mA/mm)$^b$ | 2 | 300 K | [39] |
| | Graphene/hBN/ Graphene | N, tunneling | 15 | 4 | 7 K / 300 K | [27] |
| | $Graphene/WSe_2$ | N, tunneling | 250 / 200 | 6 / 4 | 1.5 K / 300 K | [62] |
| | *Graphene/hBN/ Graphene* | N, tunneling | 26 | 2 | 2 | [63] |
| | *Graphene/hBN/ Graphene* | N, tunneling | 150 | 1.3 | 300 K | [64] |
| | *Graphene/hBN/ Graphene* | N, tunneling | 0.03 | 1.3 | 300 K | [65] |
| **Carbon Nanotubes** | Single carbon nanotube | N, charge trapping / detrapping | ~$10^5$ (10 nA per nanotube)[a] | 1.4 | 300 K | [8] |
| | Suspended single carbon nanotube | **N, Joule-heating** | ~$10^8$ (10 µA per nanotube)[a] | ~2 | 300 K | [42] |
| **Nano structures** | **CNWs** | **S, Joule-heating** | **$10^5$** | **2** | **37 K** | **This work** |

[a]we assume the effective cross-section of a molecule and a carbon nanotube 10 $nm^2$
[b]we assume the effective thickness of graphene 0.34 nm

Fig. 4(d) shows the temperature distribution of 100 microns channel at temperature 35K sitting on the substrate of 470 microns thick. The latter is placed on a stage at a temperature of 5K. The effective value of κ is enhanced from a one-dimensional model result of κ is enhanced from a one-dimensional model result $\kappa_{ox}/t_{ox}$ by a factor of $\alpha \approx 4.6$, where $\kappa_{ox}$ is the silicon substrate thermal conductivity of thickness $t_{ox}$. Therefore, our fitted value of κ=12000 W/$Km^2$ would translate to $\kappa_{ox} = \kappa t_{ox}/\ \alpha \approx 1.2$ W/Km. This value of $\kappa_{ox}$ is two orders of magnitude lower than the thermal conductivity of Si in the temperature range of interest here [66]. Therefore, interfacial thermal boundary resistance determines the thermal management of our devices. For two-dimensional graphene [67–69] lying flat on $SiO_2$ substrates, the interfacial thermal boundary conductance was reported in the range of MW/$Km^2$, which is also higher than our effective κ. The difference is due to our CNW material's morphology, shown in a revised Fig. 1b, where an effective contact area is substantially reduced from a flat 2D layer geometry.

In table 2, we compare the characteristics of our material with reported materials classes showing NDR behavior. The tunneling mechanism is primarily responsible for NDR in pioneering semiconducting materials. The switching time in tunneling devices is fast, and room temperature operation was demonstrated. However, the on-currents are typically not that high. The molecular structures demonstrate NDR due to the tunneling mechanism with high current density per molecule. However, their integration is very challenging. A revived interest took place with the discovery of NDR with compelling characteristics in oxide materials. The dominant mechanisms, electrochemical reaction and Joule heating, are typically slower than the tunneling mechanism. Layered materials recently demonstrated NDR using various mechanisms, in which the low tunneling probabilities limit current densities and integration challenges of stacking layers need to be addressed.

The CNW material reported here is easy to integrate and cheap to produce. The current densities are high, and the peak-to-valley ratio is reasonable. The microscopic mechanism for temperature dependence needs to be understood better to enable engineering activation energy in Eq. (1) to increase the operating temperature. The speed of operation is determined by the time needed to heat up and cool down the device. The switching time can be obtained from the time-dependent version of Eq. (2) with an additional term $Ch_{CNW}\frac{dT(x,t)}{dt}$ in the righthand side of Eq. (2), where $C$ is electronic specific heat. Then Eq. (2) becomes drift-diffusion equation with the effective diffusion coefficient $D = \kappa_{CNW}(T(x))/C$. Therefore, characteristic switching time can be estimated as $\tau \propto L^2/D = L^2 C/\kappa_{CNW}$. Following Ref. [70], we can estimate electronic specific heat using the graphite value $C = 13.8\,T$ μJ/(mol K) or $2.6T$ J/(m$^3$K) [71]. The time response of $\tau \propto 6$ ns can be estimated in a device with the channel length $L$=$L_c$=15 μm (see Fig. 3b) and $\kappa_{CNW} = 3$ W/mK at $T$=30 K (see Fig. 3a). This estimate implies that device operation is suitable near the microwave range (up to 0.2 GHz frequencies). Should the activation energy be engineered to enable room-temperature operation, the specific heat would increase proportionally by a factor of 10. At the same time, thermal conductivity, according to Fig. 3a, would also increase by order of magnitude, such that the time characteristics would be similar.

## IV. Conclusions

Here we present a nanostructure class of materials demonstrating strong S-type NDR effects. Carbon NanoWalls maintain high current densities, unlike tunneling-based devices with NDR characteristics. The self-consistent electrical and thermal transport simulations prove that the heat sink to the substrate and the strong temperature dependence of the low-bias resistance are the main reasons for the nonlinear electrical behavior. Our studies suggest a pathway towards optimizing and controlling the NDR regime by the heat management and fabrication process of these electrical devices based on Carbon NanoWall low-dimensional materials. The simulations of the switching time suggest GHz frequency operation is possible in ultimately scaled devices. Further studies are needed to increase the operating temperature by engineering activation energy in Carbon NanoWalls.

## Acknowledgments

S. A. E. and M. A. T. have made equal contributions to this work. V. P. gratefully acknowledges support from the Vice President for Research and Economic Development (VPRED) at the University at Buffalo and the Center for Computational Research at the University at Buffalo [72].